\documentclass[3p,twocolumn]{elsarticle}
\pdfoutput=1
\usepackage{amsmath}
\usepackage{amsfonts}
\usepackage{amssymb}
\usepackage{lineno}
\usepackage[utf8]{inputenc}
\usepackage[colorlinks=true]{hyperref}
\usepackage[nameinlink]{cleveref}

\journal{NIM A}

\begin{document}

\begin{frontmatter}

    \title{Analytical model for the uncorrelated emittance evolution of externally injected beams in plasma-based accelerators}

    \author[desy]{Alexander Aschikhin}
    \author[desy,uni-hamburg]{Timon Johannes Mehrling}
    \author[uni-hamburg]{Alberto Martinez de la Ossa}
    \author[desy]{Jens Osterhoff}

    \address[desy]{Deutsches Elektronen-Synchrotron DESY, 22607 Hamburg, Germany}
    \address[uni-hamburg]{Institut für Experimentalphysik, Universität Hamburg, 22761 Hamburg, Germany}

    \begin{abstract}    
	This article introduces an analytical formalism for the calculation of the evolution of beam moments and the transverse emittance for beams which are externally injected into plasma wakefield accelerators. This formalism is then applied to two scenarios with increasing complexity -- a single beam slice without energy gain and a single beam slice with energy gain, both propagating at a fixed co-moving position behind the driver. The obtained results are then compared to particle-in-cell (PIC) simulations as well as results obtained using an semi-analytic numerical approach (SANA) \cite{SANA}. We find excellent agreement between results from the analytical model and from SANA and PIC.
	\end{abstract}


\end{frontmatter}

\section{Introduction}

Plasma wakefield accelerators (PWFA), once a novel concept promising a significant improvement of acceleration parameters such as the energy gradient \cite{PWFA}, are now considered as a potential core component for the next generation of compact particle accelerators, following a series of successful experiments yielding highly promising results such as the energy doubling of 42 GeV electrons \cite{blumenfeld2007} or the acceleration of an electron bunch with a high energy-transfer efficiency \cite{litos2014}.
The acceleration process itself consists of a two-beam setup, involving a highly relativistic, high current \emph{drive beam} and a \emph{witness beam} propagating within a specific offset to each other in a plasma environment. 
While the former creates plasma density waves, the latter is accelerated within the resulting wakefields. 
The driver is provided by a preacceleration process ahead of the plasma cell, while the witness can be injected either externally or obtained from the electrons available within the plasma, a process known as internal injection.

A major aspect of plasma-based acceleration is the unique ability of the plasma environment to sustain accelerating gradients of 1--100 GV/m \cite{bingham2003}, which is an increase of several orders of magnitude compared to the conventional cavities used for particle acceleration today. 
Multiple design proposals exist aiming to explore this potential, suggesting applications ranging from compact X-ray sources \cite{kneip2010} to next-generation particle collider facilities \cite{seryi2009}. 
The FLASHForward project, a plasma-wakefield acceleration experiment at DESY \cite{aschikhin2015}, aims to analyze the potential of PWFA accelerated beams for free-electron laser (FEL) gain. 
The stringent requirements of FELs, however, severely limit the range of acceptable beam quality parameters, necessitating extensive studies focusing on beam quality preservation during the multiple stages involved -- from the preacceleration stage in the FLASH accelerator, to the vacuum-to-plasma transition regions as well as beam extraction after the PWFA stage. 

As an essential beam quality parameter, e.g. for FEL applications, the transverse phase space emittance received extensive attention in several works concerned with plasma-based acceleration processes \cite{michel2006,mehrling2012}. 
Furthermore, it was shown that transitions into and the propagation within multiple stages can lead to significant emittance growth if the beams are not matched and the vacuum-to-plasma and plasma-to-vacuum transitions not tapered properly \cite{mehrling2012,marsh2005,floettmann2014,dornmair2015}.

In general, an emittance increase is related to a change in the shape of the phase space volume occupied by the beam. 
This effect can be caused by multiple factors, from an off-axis injection to nonlinear transverse forces or coupling effects in the transverse-longitudinal beam particle motion. 
In the context of the blowout regime considered in this article, however, it is the mismatch between the beam and the plasma environment, together with significant correlated or uncorrelated energy spreads or variations of the longitudinal fields over the intra-bunch length, that most strongly contribute to an emittance degradation.

The extensive analysis of such a critical component and the corresponding processes typically involves the use of particle-in-cell (PIC) simulations, which offer valuable insights at the cost of time consuming computations, especially when performing parameter scans. 
And while models such as the semi-analytic numerical approach \cite{SANA} provide a significant increase in efficiency, the analytic model encompasses the advantage to not only immediately deliver the relevant parameters and their evolution, it could also offer direct insights on the impact of the different components involved in the acceleration process.

This article presents a set of analytic descriptions of the evolution of beam moments in a PWFA process for two different scenarios. 
The first scenario involves a slice of the witness beam with a specific uncorrelated energy spread propagating in a PWFA blowout regime without energy gain. 
The second scenario considers the same situation, additionally taking into account an energy gain along the propagation axis. 
Both scenarios are discussed within their respective sections, which follow the same structure -- an initial introduction of the physical and mathematical basis for the consideration, a description of the physical environment, as well as a depiction and analysis of the resulting formulas where appropriate, followed by a comparison with PIC simulations and the aforementioned SANA calculation results. 
The paper is finalized with a summary and conclusion.

\section{Scenario I -- beam slice without energy gain}

In general, the beam emittance is a six-dimensional phase space volume with a conserved density along any particle trajectory \cite{floettmann2003}. Usually, however, two-dimensional projections into orthogonal planes are considered (e.g. ${x}$-${p_x}$), occupying an area comprised of particle positions limited to a core with a given distribution. In the following section, we derive an analytic description for the evolution of such a phase space area, starting with individual particle trajectories and considering their phase space distribution to arrive at a description of their beam moments and, consequently, the beam emittance.

\subsection{Mathematical Model}

We begin by considering the differential equation for the transverse position $x$ of a single electron exhibiting no change in energy within a linearly focusing ion-channel

\begin{equation}
\label{eq:s1_1}
\frac{d^2 x}{dt^2} + \omega^2_{\beta}x = 0,
\end{equation}
with the betatron frequency $\omega_{\beta} = \omega_p / \sqrt{2 \gamma}$, the Lorentz factor $\gamma$, and where $\omega_p = \sqrt{4 \pi n_0 e^2 / m}$ is the plasma frequency, with the ambient plasma density $n_0$, the elementary charge $e$ and the electron rest mass $m$, thus constituting a harmonic oscillator. The solution for \Cref{eq:s1_1} is given by

\begin{equation}
\label{eq:s1_2}
x(t) \simeq x_0 \cos[\varphi(t)] + \frac{p_{x,0}}{m \gamma_0 \omega_{\beta,0}} \sin [\varphi(t)],
\end{equation}
with the initial position $x_0$, the initial transverse particle momentum $p_{x,0}$ as well as the initial Lorentz factor $\gamma_0$ and betatron frequency $\omega_{\beta,0}$. 
The phase advance, included as the argument in the trigonometric functions, is defined as ${\varphi (t) = \int \omega_{\beta}}$, with ${\varphi (t) = \omega_{\beta, 0} t}$ in this consideration.

In the following, we transition away from a single-particle picture towards a statistical approach involving collective beam slice moments in order to arrive at an analytic formulation of the normalized transverse phase space emittance \cite{floettmann2003}

\begin{equation}
\label{eq:s1_3}
    \epsilon_n = \frac{1}{m_e c}\sqrt{\left\langle x^2 \right\rangle \left\langle p^2_x \right\rangle - \left\langle x p_x \right\rangle^2},
\end{equation}
with the phase space beam moments $\left\langle x^2 \right\rangle$, $\left\langle p_x^2 \right\rangle$, $\left\langle x p_x\right\rangle^2$. We chose an ansatz to evaluate \Cref{eq:s1_2} which relies on a separable beam distribution function (as outlined in \cite{SANA}). 
We assume that the beam particle slice possesses an initial phase space distribution $f_0 (x_0, p_0, \gamma_0)$ (with the normalization $\int f_0 d x_{x,0} d p_{x,0} d \gamma_0 = 1$), where the initial transverse position $x_0$ and the initial transverse momentum $p_{x,0}$ are not correlated with the energy. 
This means that the beam distribution is separable ${f_0 = f_{\perp} (x_0, p_{x,0}) f_{\gamma} (\gamma_0)}$, thus allowing for the retrieval of the phase space moments using

\begin{align}
\langle x^2 \rangle (t) &= \int_{-\infty}^{\infty} (x^2(t)) f_0 d x_0 d p_{x,0} d \gamma_0 \\
\langle p_x^2 \rangle (t) &= \int_{-\infty}^{\infty} (p_x^2(t)) f_0 d x_0 d p_{x,0} d \gamma_0 \\
\langle x p_x \rangle (t) &= \int_{-\infty}^{\infty} \left( x (t) p_x (t) \right) f_0 d x_0 d p_{x,0} d  \gamma_0,
\end{align}
with an arbitrary $f_{\perp} (x_0, p_{x,0})$ only stipulating ${f_{\perp} = 0}$ outside of the ion-channel and the energy distribution assumed to follow a Gaussian form with ${f_{\gamma} = (\sqrt{2 \pi} \sigma_{\gamma})^{-1} \exp (-\delta \gamma^2 / 2 \sigma_{\gamma}^2)}$, where ${\delta \gamma = \gamma - \overline{\gamma}}$, describes a deviation of individual particles from the mean slice energy $\overline{\gamma}$. 

Assuming a small relative energy deviation of the electrons, ${\left|\delta\gamma/\overline{\gamma}\right| \ll 1}$, allows for an approximation of the betatron frequency ${\omega_{\beta} \simeq \overline{\omega_{\beta}}} \left( 1 - \delta\gamma/2 \overline{\gamma} \right)$, with the mean betatron frequency ${\overline{\omega_{\beta}} = \omega_p/\sqrt{2 \overline{\gamma}}}$. 
Since we ignore energy variations in this scenario, the betatron frequency remains constant ${\omega_{\beta} = \omega_{\beta,0}}$, making it possible to provide the phase advance as ${\varphi(t) = \overline{\omega_{\beta,0}} \left( 1 - \delta \gamma / 2 \overline{\gamma_0} \right) t}$. 
Using this expression, together with the original solution for the individual particle position, \Cref{eq:s1_2} as well as ${p_{x} (t) = m \gamma dx/dt}$, we can obtain the transverse beam moments and thus the transverse phase space emittance, which can be written in the following analytic expression

\begin{align}
\label{eq:s1_4}
\begin{split}
\epsilon_{n}^2 (t) = \frac{1}{4} &\left( \left(\overline{\gamma_0} \overline{k_{\beta}}\right)^2 \left\langle x_0^2\right\rangle^2
+ \frac{1}{\left(\overline{\gamma_{0}} \overline{k_{\beta}}\right)^2} \left\langle u_{x,0}^2\right\rangle^2 \right) \\
&\times \left(1 - e^{- b t^2 }\right) \\
&+ \frac{1}{2} \left\langle x_0^2\right\rangle \left\langle u_{x,0}^2\right\rangle 
\left(1 + e^{- b t^2 }\right) \\
&- \left\langle x_0 u_{x,0}\right\rangle^2  
e^{- b t^2 },
\end{split}
\end{align}
with the momentum given in ${u_x = p_x / m_e c}$, the mean betatron oscillation wave number ${\overline{k_{\beta}} = \overline{\omega_{\beta}} / c}$, together with a growth factor ${b = \overline{\omega_{\beta}}^2 \Delta \gamma^2}$ (using the common depiction of the energy spread ${\Delta \gamma = \sigma_{\gamma} / \gamma}$).
It is straightforward to recover the initial normalized phase space emittance ${\epsilon_0 = \sqrt{\left\langle x_0^2 \right\rangle \left\langle u^2_{x,0} \right\rangle - \left\langle x_0 u_{x,0} \right\rangle^2}}$ of the slice, by setting ${t=0}$.
It can also be observed that the subsequent time-dependent change in emittance is driven by the exponential terms and thus, through the growth factor, the initial energy spread of the slice. 
This betatron-phase mixing effect, owing to the finite energy spread in the slice and the corresponding energy-dependent oscillations of the electrons, is an example of the so-called betatron decoherence.
Its influence on the development of emittance growth is reflected in the prominent role of the betatron wave number (or frequency) in the first term.
It can be observed from \Cref{eq:s1_4} that the emittance growth reaches a saturation point once the contribution from the exponential term is sufficiently small, providing a time scale for the decoherence as ${t_d \gg b^{-1/2} = 1/\left( \Delta \gamma \overline{\omega_{\beta}} \right)}$.
Assuming ${t \rightarrow \infty}$, we can then derive an expression for the final beam emittance once full decoherence is reached,
\begin{equation}
\label{eq:s1_5}
\epsilon_{n}^2 = 
\frac{\left(\overline{\gamma_0} \overline{k_{\beta}} \right)^2}{4}
\left\langle x_{0}^2\right\rangle^2
+ \frac{\left\langle u_{x,0}^2\right\rangle^2}{4 \left(\overline{\gamma_0} \overline{k_{\beta}} \right)^2}
+ \frac{1}{2} \left\langle x_{0}^2\right\rangle \left\langle u_{x,0}^2\right\rangle.
\end{equation}
It is relevant to note that the energy spread of the beam slice does not play a role in this expression (reproducing previous results, see \cite{mehrling2012}).
While it is the driving factor behind the betatron decoherence and determines the time-scale of its progression, the final emittance is dictated by the initial beam parameters.
This means that a beam not properly matched to the intrinsic betatron motion in the plasma will exhibit emittance growth \cite{mehrling2012}. 
This behavior can be avoided if matching conditions are met, avoiding beam quality degradation. 
These conditions can be translated into quantities relevant for this formulation as ${\left\langle x u_x \right\rangle_m = 0}$, ${\overline{k_{\beta}} \left\langle x^2 \right\rangle_m = \epsilon_0 / \overline{\gamma_0}}$, ${\left\langle u_x^2 \right\rangle_m / \overline{k_{\beta}} = \epsilon_0  \overline{\gamma_0}}$.
Together with these expressions for matched beam moments and \Cref{eq:s1_4}, we can find an emittance growth factor,
\begin{equation}
\label{eq:s1_6}
\begin{split}
    \left(\frac{\epsilon_{n}(t)}{\epsilon_0}\right)^2
    = &\frac{1}{4}
    \left( 
    \left(
    \frac{ \left\langle x_0^2\right\rangle}{\left\langle x^2\right\rangle_m}
    \right)^2 
    + \left(
    \frac{\left\langle u_{x,0}^2\right\rangle}{\left\langle u_{x}^2\right\rangle_m}
    \right)^2
    \right) \\
    &\times \left(1 - e^{- b t^2 }\right)\\
    &+ \frac{1}{2} 
    \frac{\left\langle x_0^2\right\rangle}{\left\langle x^2\right\rangle_m}
    \frac{\left\langle u_{x,0}^2\right\rangle}{\left\langle u_{x}^2\right\rangle_m}
    \left(1 + e^{- b t^2 }\right) \\
    &- \frac{\left\langle x_0 u_{x,0}\right\rangle^2}{\epsilon_0^2} 
    e^{- b t^2 }.
\end{split}
\end{equation}
It is evident from \Cref{eq:s1_6} that following the matching requirements for the initial values results in a growth factor of one, equivalent to a preservation of the initial transverse phase space emittance.

\subsection{Physical Studies}

In order to benchmark the analytic description presented above, we run a Particle-in-Cell (PIC) simulation, using the 3D quasi-static code HiPACE \cite{hipace2014}.
The blowout regime, which allows us to assume no radial dependence of the longitudinal wakefield and thus decouple the radial and longitudinal phase space distributions, was established using a Gaussian drive beam with a peak current ${I_b = 3 \mathrm{kA}}$, total charge ${Q_b = 240 \mathrm{pC}}$, mean energy ${\overline{\gamma_0} = 2000}$, energy spread ${\sigma_{\gamma}/\overline{\gamma_0} = 0.1 \%}$ and a transverse phase space emittance ${\epsilon_{n} = 2.0 \mathrm{\mu m}}$, moving through a homogeneous plasma of constant density (i.e. without a tapered vacuum-to-plasma transition) of ${n_p = 10^{23} \mathrm{m}^{-3}}$ (with the local peak density of the beam ${n_b/n_p = 28.5 \gg 1.0}$).
The witness beam was modeled as a slice of macroparticles with a mean energy ${\overline{\gamma_0} = 2000}$, an energy spread of ${\sigma_{\gamma} / \overline{\gamma_0} = 10\%}$, a transverse phase space emittance of ${\epsilon_{0} = 2.0 \mathrm{\mu m}}$ and initial root mean square (rms) beam moments ${\sigma_{x,0} = \sqrt{\left\langle x^2_0 \right\rangle } = 5 \mathrm{\mu m}}$, ${\left\langle x_0 u_{x,0} \right\rangle = 0}$, and the momentum spread thus given by ${\sigma_{p_x,0} = \sqrt{\left\langle u^2_{x,0} \right\rangle} = \epsilon_0/\sqrt{\left\langle x^2_0\right\rangle}}$.
The particular values reflect a commonly encountered parameter range, for example at the FLASHForward experiment \cite{aschikhin2015}.

Following the requirements of the analytic model, the witness slice was placed behind the driver at the zero-crossing of the electric field, avoiding changes in its energy as much as possible.
An additional benchmark was provided by an implementation of the semi-analytic numerical approach (SANA), using a calculation based on the parameters provided for the PIC simulation.

\begin{figure}[!htbp]
\includegraphics[width=1.0\columnwidth]{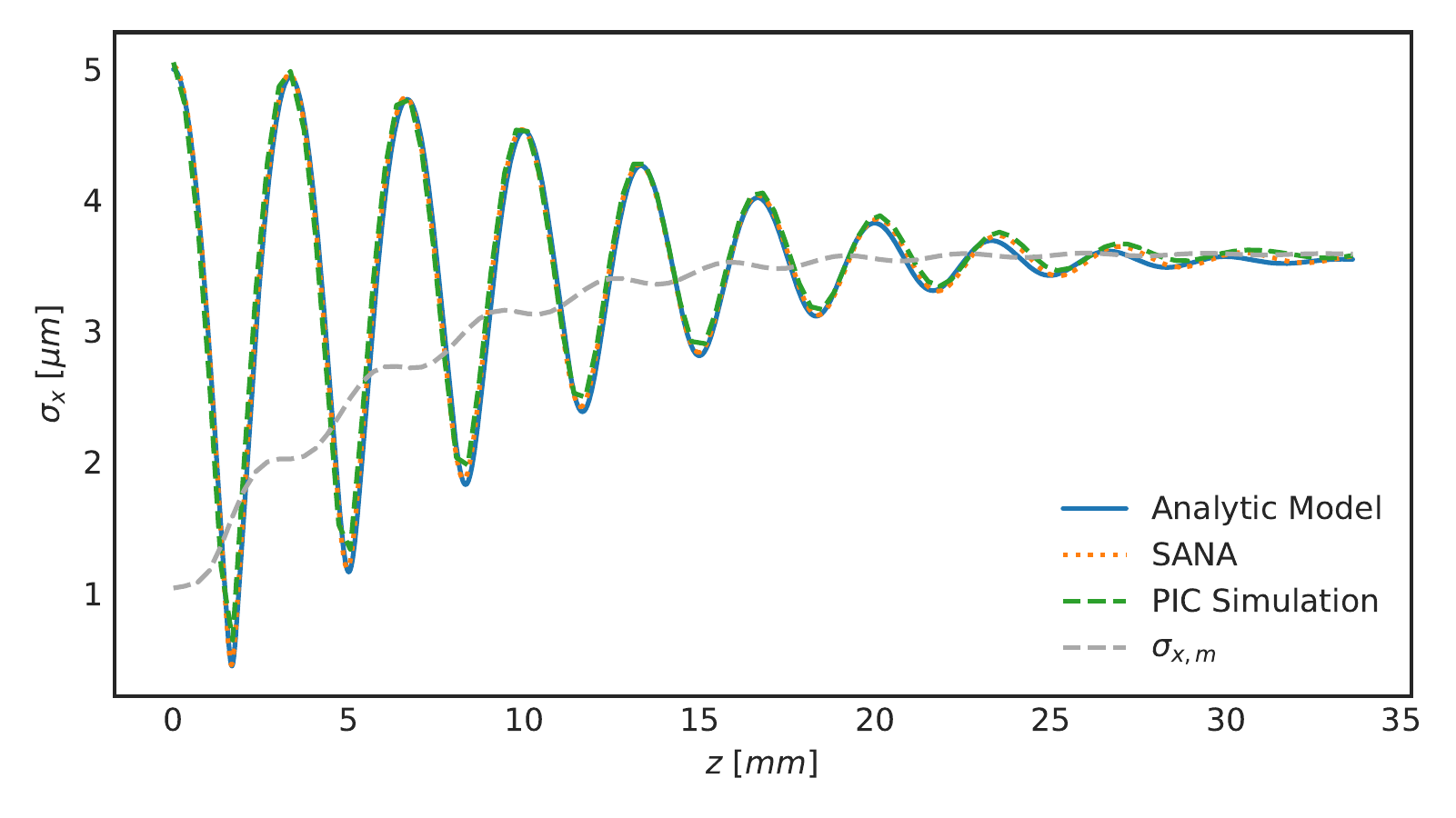}
\caption{Evolution of the beam size ${\sigma_x}$ and the instantaneously matched parameter ${\sigma_{x,m}}$.}
\label{fig:s1_1}
\end{figure}
\begin{figure}[!htbp]
\includegraphics[width=1.0\columnwidth]{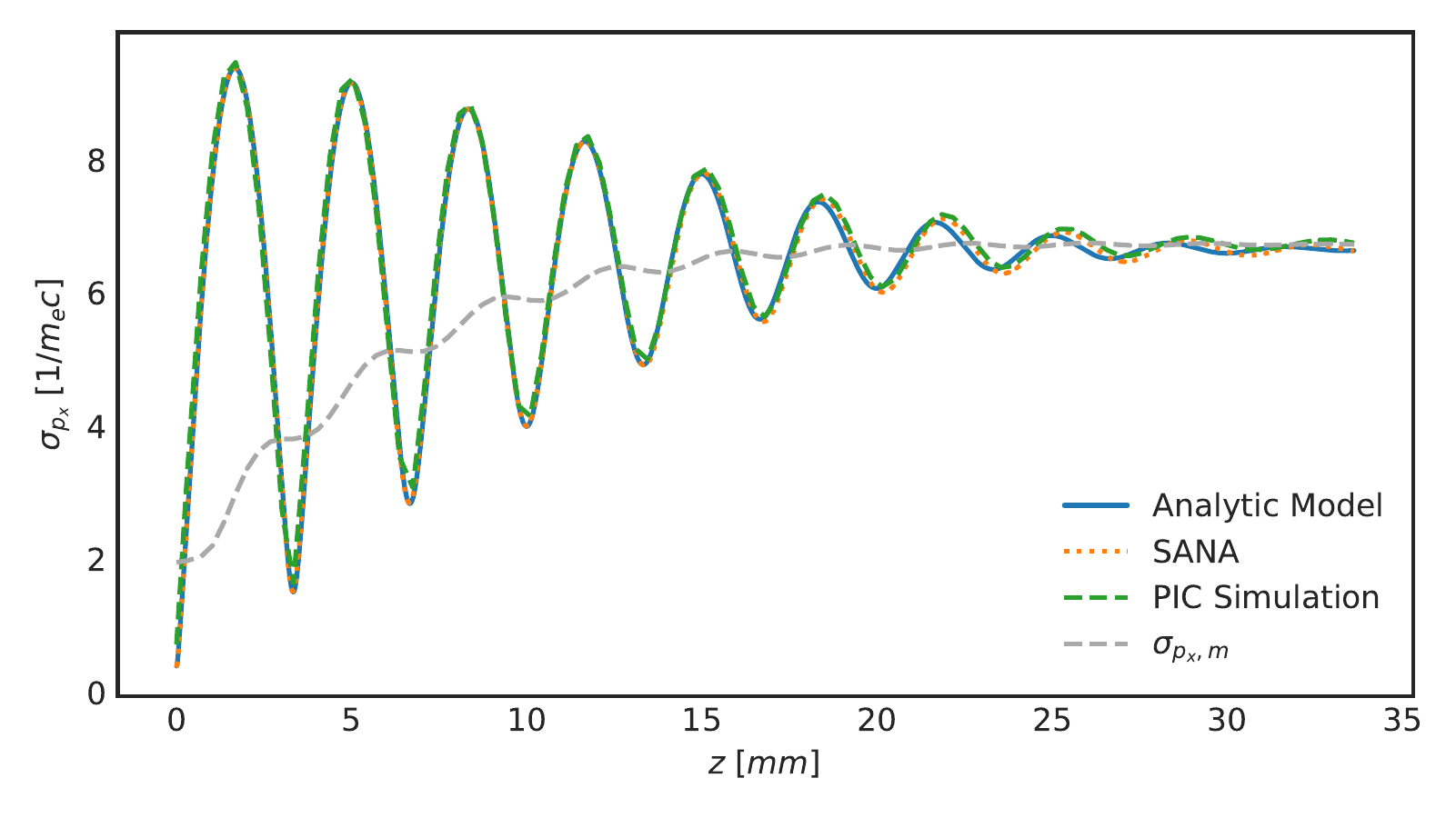}
\caption{Evolution of the rms beam moment ${\sigma_{p_x}}$ and the instantaneously matched parameter ${\sigma_{p_x,m}}$.}
\label{fig:s1_2}
\end{figure}
\begin{figure}[!htbp]
\includegraphics[width=1.0\columnwidth]{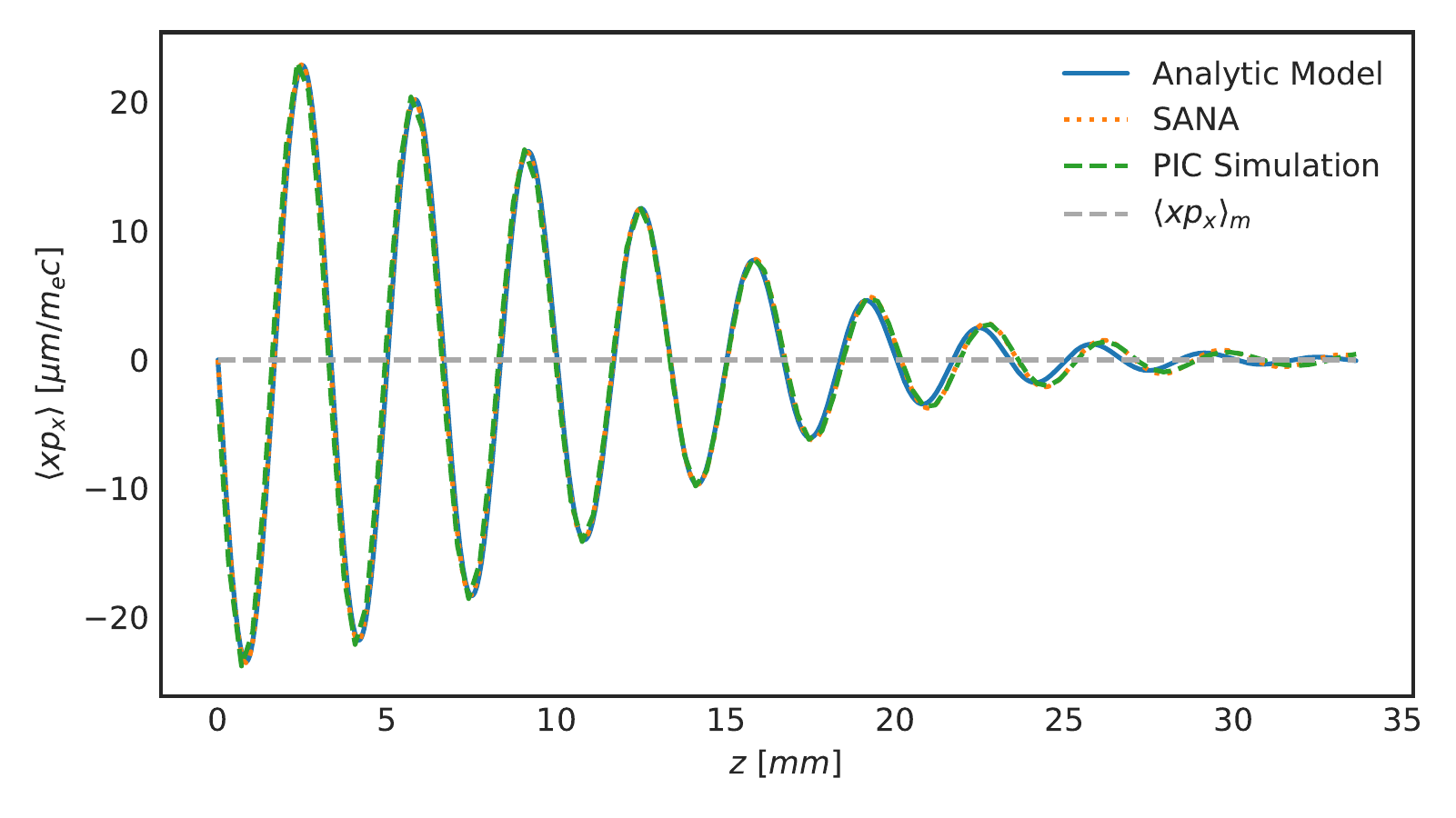}
\caption{Evolution of the correlation beam moment ${\left\langle x \cdot p_x\right\rangle}$ and the matched parameter ${\left\langle x \cdot p_x\right\rangle}_m$.}
\label{fig:s1_3}
\end{figure}
\begin{figure}[!htbp]
\includegraphics[width=1.0\columnwidth]{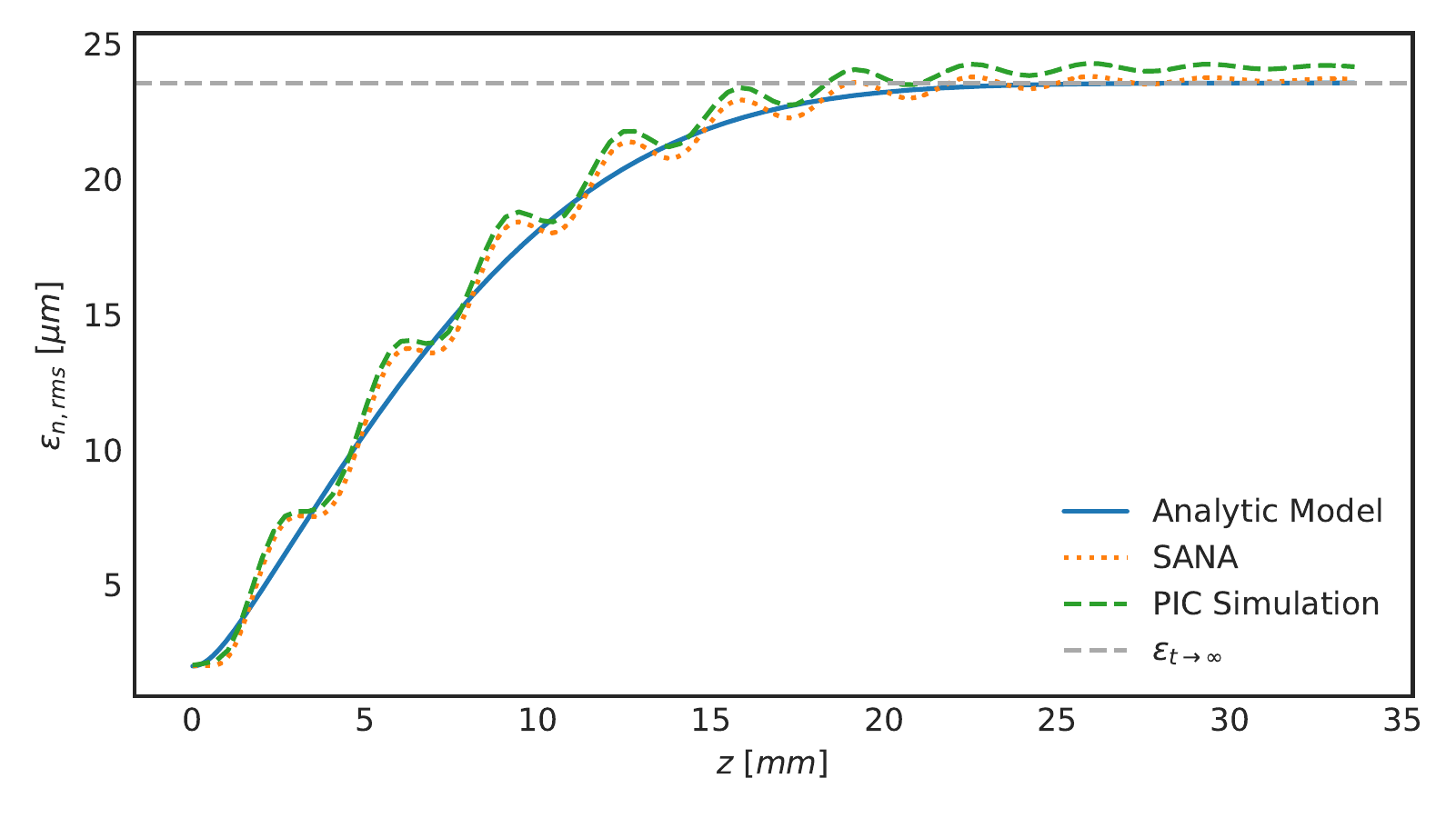}
\caption{Evolution of the transverse phase space emittance $\epsilon_{n}$ together with the value for the final emittance obtained from the analytic model according to \Cref{eq:s1_5}, and plotted as an upper boundary.}
\label{fig:s1_4}
\end{figure}

The results are provided in \Crefrange{fig:s1_1}{fig:s1_4} depicted in SI units using a notation for the rms beam moments where ${\sigma_x = \sqrt{\left\langle x^2 \right\rangle}}$, ${\sigma_{p_x} = \sqrt{\left\langle p^2_{x} \right\rangle} }$.
Additionally, we provide beam parameters dynamically matched to the current emittance and energy, given as ${\sigma_{x,m} = \sqrt{\epsilon/k_p  \sqrt{2/\overline{\gamma}}}}$ and ${\sigma_{p_x,m} = \sqrt{\epsilon k_p \sqrt{\overline{\gamma}/2}}}$. 
This definition is reused for the physical studies of the next section as well, when both the emittance and the energy vary over the simulation length.
Because of the mismatched initial beam parameters, the emittance of the slice grows significantly during the propagation, approaching the calculated final emittance value.
As mentioned above, this is due to the energy-dependence of the betatron frequency, causing its decoherence and thus an increase in the emittance of the slice.
The decoherence can be observed as a damping effect on the beam moment oscillations, which ultimately approach their matched values once complete decoherence is reached.
Using the growth factor derived above, we can provide a time scale for full decoherence for the given parameters as ${t_d \gg b^{-1/2} = 1/\left( \Delta \gamma \overline{\omega_{\beta}} \right)}$, transferring it into a distance to better match the plots, ${z_d \gg c/\left( \Delta \gamma \overline{\omega_{\beta}} \right) \approx 10.6 \mathrm{mm}}$.
Over these lengths, no significant drive beam head erosion can be observed in the simulations, allowing us to assume ${E_z}$ to be constant for the simulation time and distance.
Since we are concerned with a single slice of electrons without acceleration, we can ignore correlated emittance growth due to an energy chirp for a witness beam of finite length (see \cite{SANA, mehrling2012}) and focus on the uncorrelated emittance growth.
However, while the macroparticle slice in the PIC simulations was chosen to be as thin as possible in the longitudinal direction (${\sigma_{\zeta} \ll k^{-1}_{p,0}}$), it nevertheless samples the electric field variation over its length and around the zero-crossing of $E_z$, thus deviating from the assumption of no energy variation for the witness beam, potentially contributing to a slightly higher emittance value than the one we calculated using the analytic model.
Otherwise, we find excellent agreement between the analytic model and the numerical results.

\section{Scenario II -- beam slice with energy gain}

\subsection{Mathematical Model}

The second scenario exhibits an increase in complexity, by incorporating a variation in energy, while keeping the single slice picture, again allowing to ignore the correlated emittance growth. 
In more physical terms, translated into a PIC simulation setup, it can be thought of as a thin layer of macroparticles with a transverse distribution following a driver in the blowout regime at an offset where a non-negligible longitudinal electric field is providing an accelerating gradient. 

The change in energy is reflected in an updated differential equation for the transverse position of a single particle,

\begin{equation}
\label{eq:s2_1}
    \frac{d^2 x}{dt^2} + \frac{\dot{\gamma}}{\gamma} \frac{dx}{dt} + \omega^2_{\beta} (t) x = 0,
\end{equation}
with ${\dot{\gamma} = d\gamma/dt}$ and where the acceleration of the electron leads to a damping of the particle oscillation through the term ${\dot{\gamma}/\gamma}$ (consequently, a loss in energy would result in an amplification of the oscillation amplitude). \Cref{eq:s2_1} has the solution

\begin{equation}
\label{eq:s2_2}
x(t) \simeq x_0 A(t) \cos[\varphi(t)] + \frac{p_{x,0}}{m_e \gamma_0 \omega_{\beta,0}} A(t) \sin [\varphi(t)],
\end{equation}
with the newly introduced amplitude term ${A(t) = [\gamma_0 / \gamma(t)]^{1/4}}$. 
To arrive at this description, terms of the form ${\left| \dot{\gamma} \dot{A} / (\dot{\varphi} \gamma A) \right| \ll 1 }$, ${\left| \ddot{A} / \left( \dot{\varphi}^2 A \right) \right| \ll 1}$ and ${\left| \dot{\gamma} / 4 \gamma_0 \omega_{\beta, 0} \right| \ll 1}$ were dropped, resulting from the the assumption that energy variations happen on time scales much longer than the betatron oscillations (compare \cite{mehrling2017}). 
The energy of a single electron is given as ${\gamma (t) = \overline{\gamma_0} + \mathcal{E} t + \delta \gamma}$, again with the initial mean energy ${\overline{\gamma_0}}$, the uncorrelated energy spread $\delta\gamma$ and a linear term incorporating a change in energy, where ${\mathcal{E} = -e E_z /m_e c}$ and ${E_z = E_z ({\zeta})}$ is the longitudinal electric field.
The variation in energy also means a time dependent electron betatron oscillation term ${\omega_{\beta} (t)}$ and the resulting phase advance

\begin{equation}
 \varphi (t) = \int \omega_{\beta} \mathrm{d} t
= \overline{\varphi } \left(1 - 
\frac{\delta \gamma}{2 \overline{\gamma _0}} 
\frac{\overline{\omega _{\beta }}}{ \overline{\omega _{\beta ,0}}}\right),
\end{equation}
with the mean phase advance ${\overline{\varphi } = 2 \left(\overline{\omega_{\beta ,0}}/\overline{\omega _{\beta }} - 1\right) / \epsilon }$, ${\overline{\omega_{\beta}} (t) = \overline{\omega_{\beta ,0}} / \sqrt{\epsilon  \overline{\omega _{\beta ,0}} t + 1}}$ and the finite relative energy change per betatron cycle ${\epsilon = - \sqrt{2 / \overline{\gamma_0}} E_z/E_0}$, with the cold nonrelativistic wavebreaking field ${E_0 = \omega_p m c / e}$ \cite{dawson1959}.
Using the phase advance description together with the term for energy development allows for the evaluation of the beam moment equations and a calculation of the emittance.
Unfortunately, the resulting formula cannot be provided in a concise form within the format of this publication.
The results from its application to an acceleration setup, however, are presented in the following section.

\subsection{Physical Studies}

The setup chosen for benchmarking the model developed following the restrictions imposed by the second scenario is similar to the one used for the first scenario -- a beam slice with an uncorrelated energy spread following a driver at an offset, this time with a non-zero longitudinal electric field resulting in an energy gain. 
With the other values such as the transverse beam moments kept the same, the slice placement was chosen so that $E_z (\zeta) \approx 0.3 \cdot E_0$. 

\begin{figure}[!htbp]
\includegraphics[width=1.0\columnwidth]{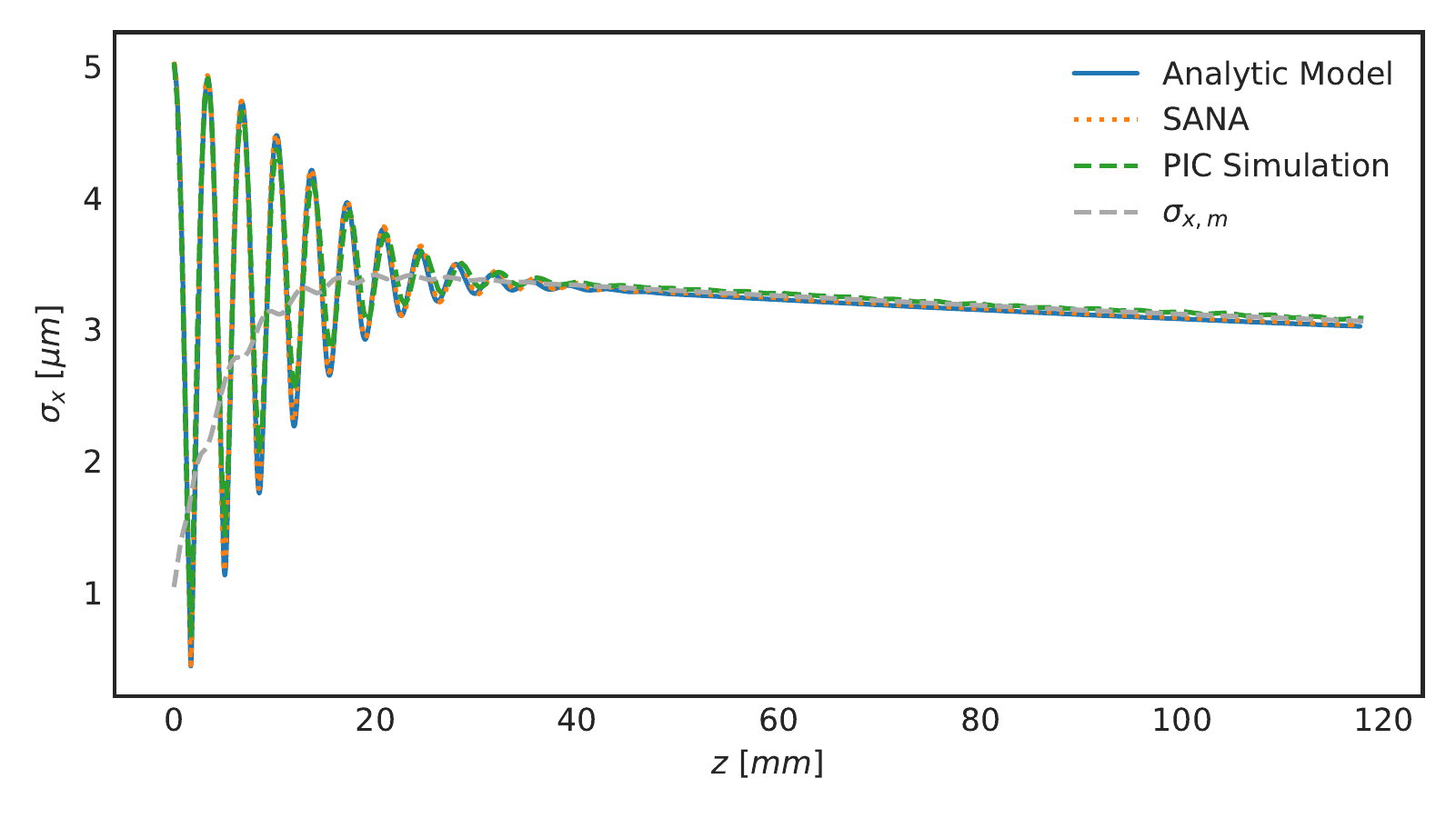}
\caption{Evolution of the beam size ${\sigma_x}$ and the instantaneously matched parameter ${\sigma_{x,m}}$.}
\label{fig:s2_1}
\end{figure}
\begin{figure}[!htbp]
\includegraphics[width=1.0\columnwidth]{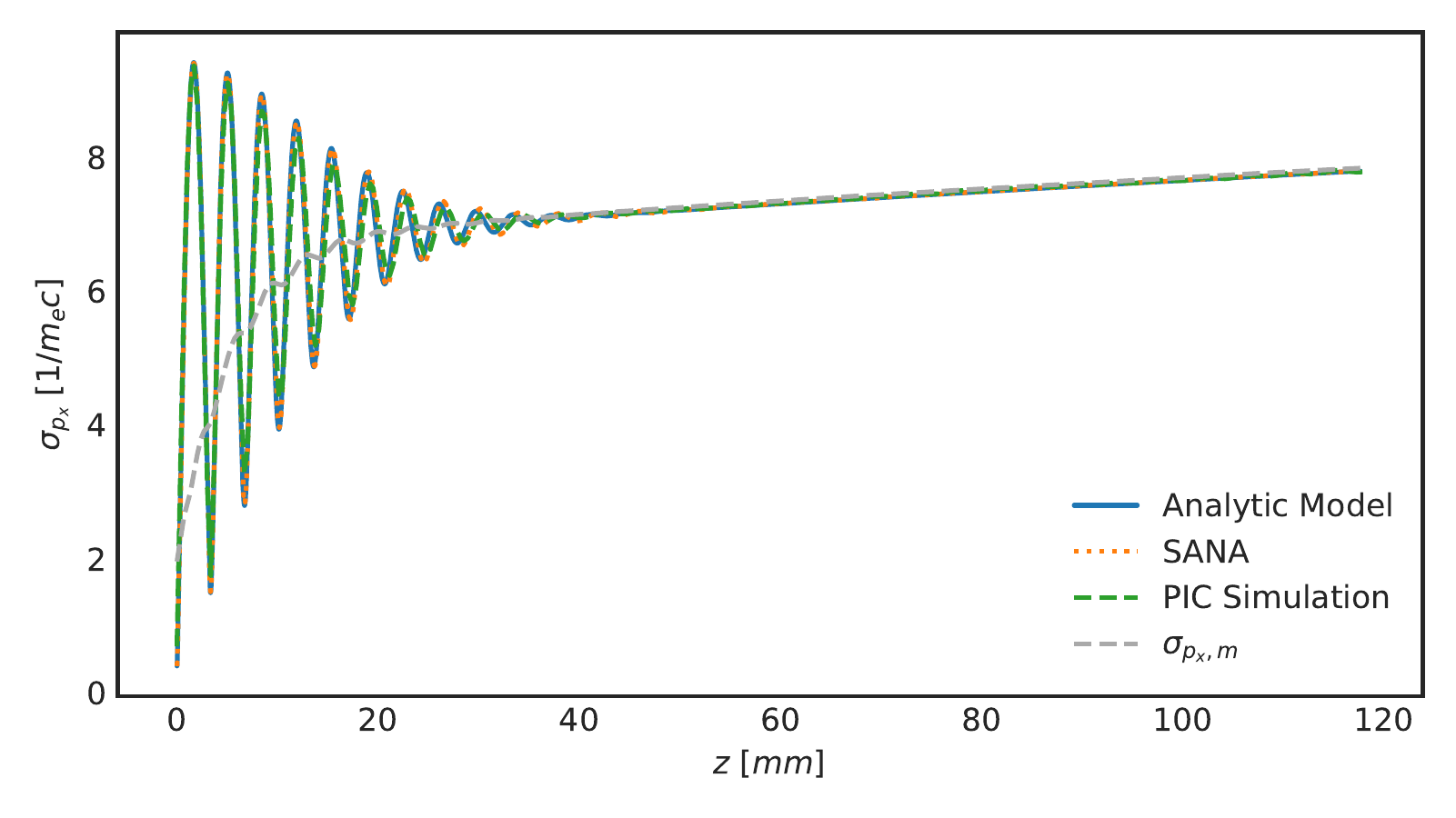}
\caption{Evolution the rms beam moment ${\sigma_{p_x}}$ and the instantaneously matched parameter ${\sigma_{p_x,m}}$.}
\label{fig:s2_2}
\end{figure}

\begin{figure}[!htbp]
\includegraphics[width=1.0\columnwidth]{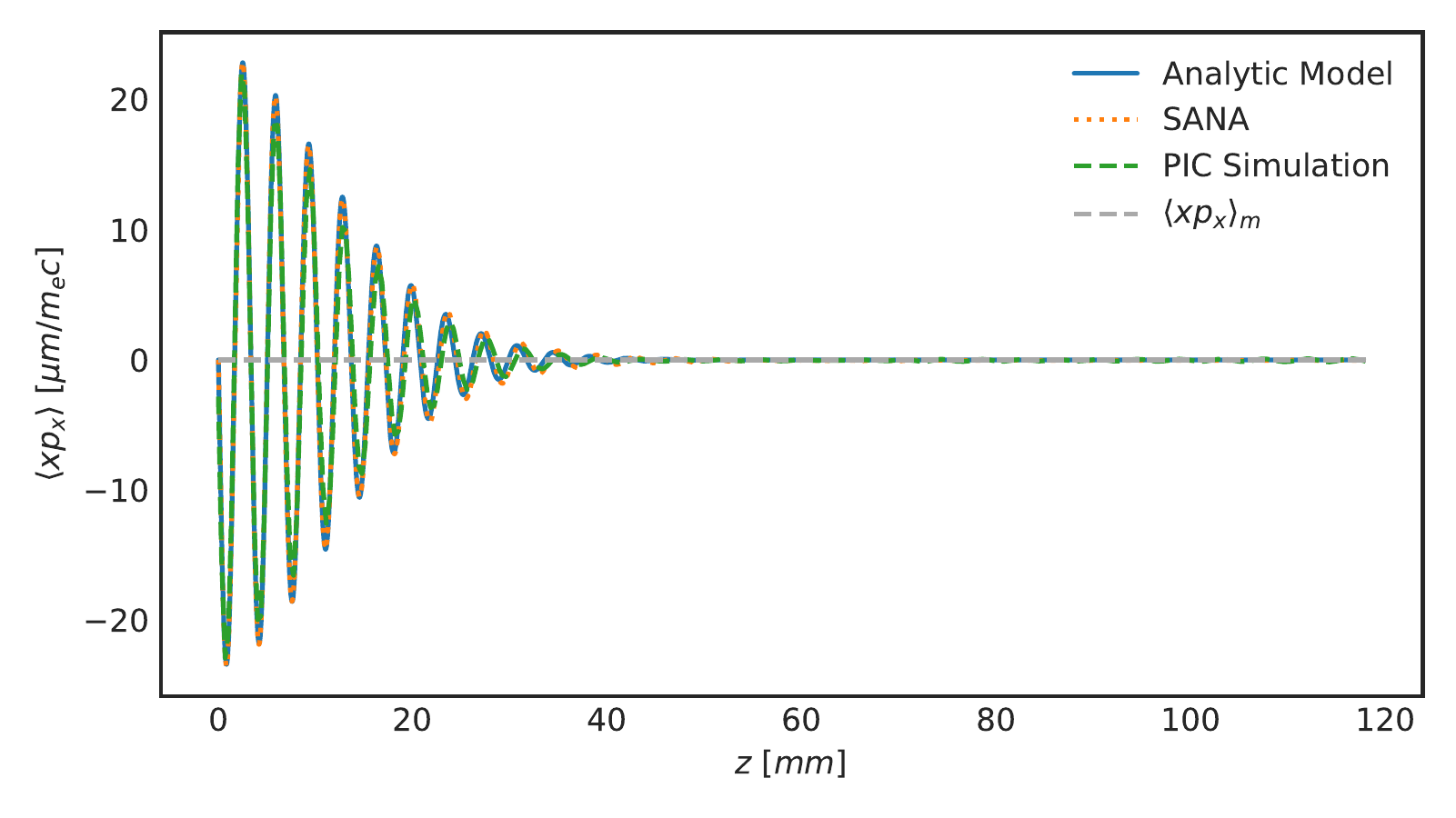}
\caption{Evolution of the correlation beam moment ${\left\langle x \cdot p_x\right\rangle}$ and the matched parameter ${\left\langle x \cdot p_x\right\rangle}_m$.}
\label{fig:s2_3}
\end{figure}

\begin{figure}[!htbp]
\includegraphics[width=1.0\columnwidth]{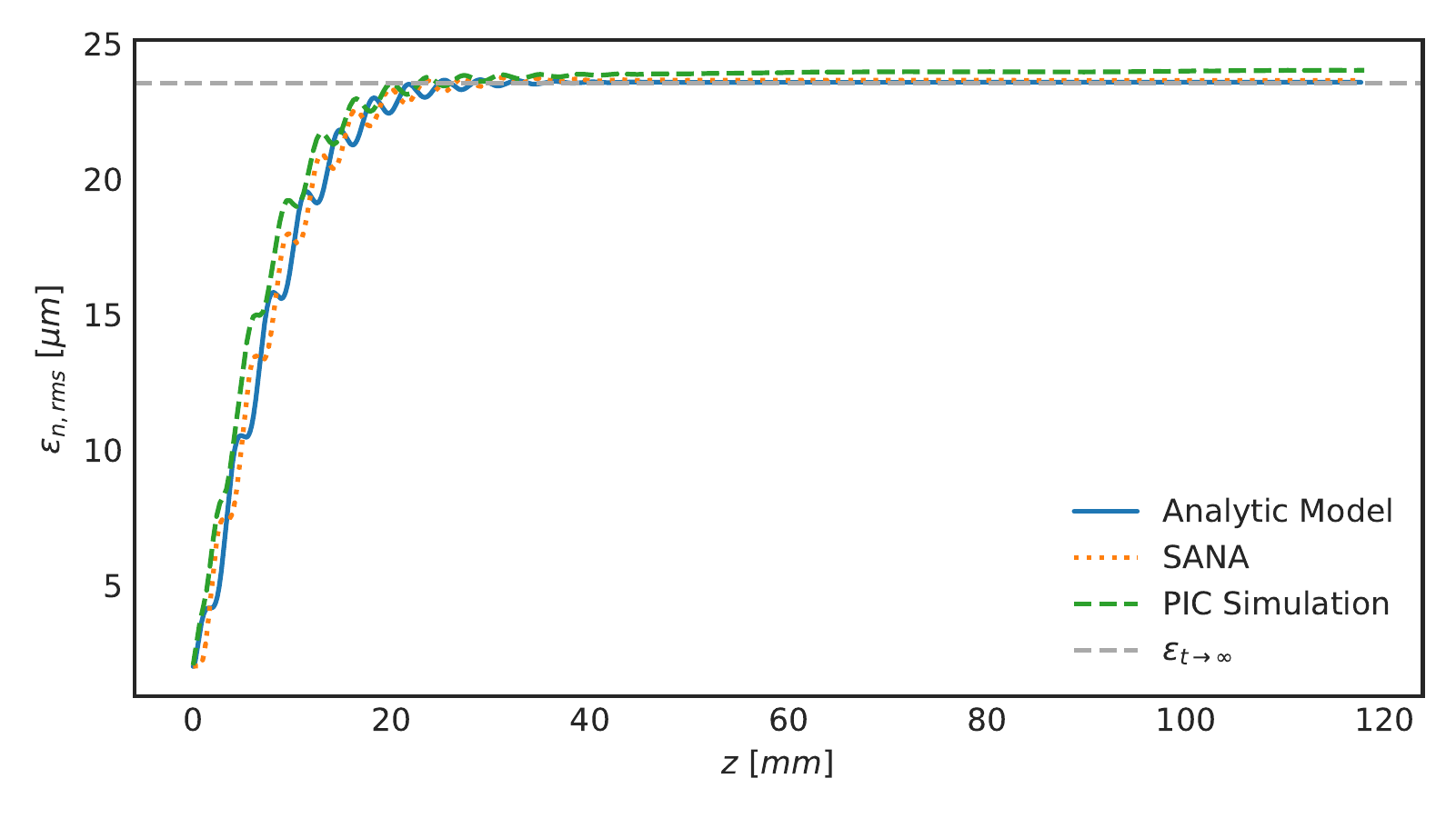}
\caption{Evolution of the transverse phase space emittance $\epsilon_{n}$, together with a final emittance value obtained from the analytic model according to \Cref{eq:s1_5}.}
\label{fig:s2_4}
\end{figure}

\Crefrange{fig:s2_1}{fig:s2_4} show the evolution of the respective beam parameters with energy gain, provided with the matched parameters for the varying emittance and energy as introduced in the physical studies section of the first scenario.
Again, we observe oscillations of the beam moments for the chosen, mismatched initial transverse beam parameters, eventually approaching the matched parameters after full decoherence due to the energy variation in the slice.
The subsequent propagation within the plasma shows a variation in the rms beam moments ${\sigma_x}$ and ${\sigma_{p_x}}$, resulting from the energy-dependency of the matched parameters ${\sigma_{x,m}}$ and ${\sigma_{p_x,m}}$ (or ${\left\langle x^2 \right\rangle}_m$ and ${\left\langle p_x^2 \right\rangle}_m$) -- originating from the amplitude term in \Cref{eq:s2_2}.


At the same time, the energy spread drives a significant increase in emittance until the matched values are reached.
Since the mechanism for the emittance growth is the decoherence effect caused by the energy-dependent oscillations of particles, we expect the subsequent acceleration within the plasma to have no effect on the development of the normalized emittance once the matched parameters are reached (see \cite{michel2006,mehrling2012}).
Thus, we plotted the final value of the emittance obtained from \Cref{eq:s1_5} to observe that it can indeed be seen as a target value for the evolution of the emittance over the propagation length.
Additionally, it should be noted that the energy spread used for benchmarking our results is assumed to be much higher than expected for externally injected witness beams in proposed PWFA experiments \cite{aschikhin2015}, with typical values of $\Delta \overline{\gamma_0} = 0.1 \%$. 
Using such an energy spread while keeping all the other parameters to recalculate the decoherence time scale using the growth factor $b$ results in a distance on the order of ${z_d \gg c/\left( \Delta \gamma \overline{\omega_{\beta}} \right) \approx 1.06 \mathrm{m}}$.
While beyond the plasma target dimensions proposed, it is nevertheless on the order of the plasma cell length and can thus play a role during the acceleration process within. 

The emittance plots again show a slightly higher value for the PIC simulation, due to the finite beam length and the resulting contribution from correlated emittance growth, mirroring the observation from the first scenario.
Apart from these minor deviations, we find a very high accuracy of the analytic model when describing the beam moment and the emittance developments.

\section{Summary and Conclusion}

We present the development of an analytical model for the calculation of the evolution of transverse beam moments and the normalized phase space emittance, together with an investigation of the uncorrelated emittance growth of externally injected beams in plasma wakefield accelerators.
The models allow for an immediate evaluation not only of initial beam parameters with respect to their matching conditions, but also of their development over an acceleration length within a section of homogeneous plasma, providing important information such as typical length scales for emittance growth and final emittance values. 
The validity of both models is presented by benchmarking it against results obtained from two different approaches -- a standard Particle-in-Cell simulation following the scenario restrictions as close as possible, together with the semi-analytic numerical approach (SANA). 
We find excellent agreement between the analytical model for the uncorrelated emittance evolution and the numerical approaches.

\section{Acknowledgments}

We would like to thank C. Schroeder and C. Benedetti for their contributions to the development of HiPACE.
We acknowledge the use of the High-Performance Cluster (Maxwell) at DESY. 
T.~J.M acknowledges the support by the Deutscher Akademischer Austauschdienst
(German Academic Exchange Service) with funds from the Bundesministerium für Bildung und Forschung and the
Marie Sklodowska Curie Actions of the EU’s FP7 under REA grant no. 605728 (P.R.I.M.E.). 
A.~M. would like to thank the Germany Postdoctoral Exchange Program and the Helmholtz Virtual
Institute VH-VI-503, for financial support.

\section{References}

\bibliographystyle{unsrt}
\bibliography{eaac17}

\end{document}